# Scientific comparison of Mozart and Salieri

Mikhail Simkin

"Do you feel quite sure of that, Dorian?"

"Quite sure."

"Ah! then it must be an illusion. The things one feels absolutely certain about are never true. That is the fatality of faith, and the lesson of romance. How grave you are! Don't be so serious. What have you or I to do with the superstitions of our age? … Play me something. Play me a nocturne, Dorian ..."

    -- Oscar Wilde, The Picture of Dorian Gray

Do you feel quite sure that Mozart was a genius and Salieri an envious mediocrity, and that their music cannot be even compared?

To check whether this is an illusion I wrote the "Mozart or Salieri?" quiz [1].  It consists of ten one-minute audio clips. Some of them are from Mozart, other – from Salieri.  The takers are to determine the composer. Though I selected some of the most acclaimed Mozart's music for the quiz, people had great difficulties doing it.  "I'm a Mozart fan and I still got 60%," – wrote one of quiz-takers.

The distribution of the scores received by over eleven thousand quiz-takers[1] is shown in Figure 1. The average score is 6.10 out of 10 or 61.0% correct, which is not much better than random guessing.

---

[1] There were almost thirty thousand test results in the database. However a number of people took several shots at the quiz.  I cleaned the data from the second-attempt scores by selecting only the first score from each IP address. Next, I cleaned the data from the results, where one or more questions were skipped. I noticed some quiz takers answered the questions without listening to the clips.  So I eliminated the results were I could not verify that all clips were downloaded from the same IP address before the test was submitted.

Our respondents failed the test. But could this be because they are a bunch of vulgar in taste philistines? Fortunately, the quizzing software [2] records IP address, from which one can infer the location of taker's computer. Thus, I was able to select the test scores received by people who downloaded the quiz from elite locations. For the analysis I chose Ivy League schools and Oxbridge (if not for any other reason, than because I did time in those places). The average elite score is 6.25 out of 10 or 62.5% correct.

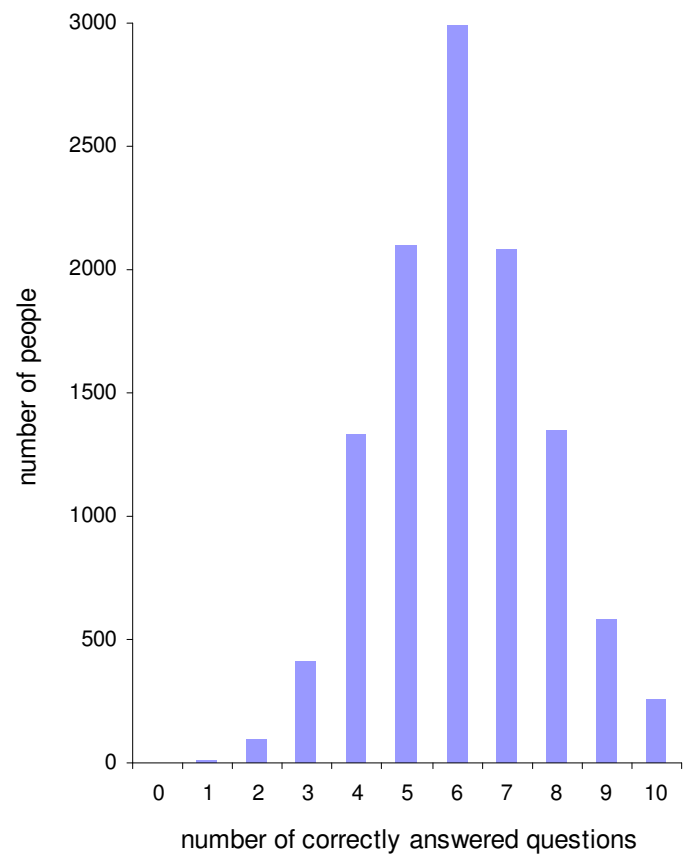

**Figure 1.** The histogram of the test scores, earned by 11,207 quiz-takers. The average score is 6.10 out of 10 or 61.0% correct.

Table 1 shows the distribution of 40 chosen quiz-takers by elite universities. The elite score, 6.25, is a bit higher than the crowd score, 6.10, but the difference is less than the standard error of elite score, which is 0.27. We see that there is no statistically significant difference between the elite and the crowd.

**Table 1** The distribution of the chosen quiz-takers by elite universities.

| Elite School | Number of quiz-takers | Score | | |
|---|---|---|---|---|
| | | min | max | average |
| Brown | 2 | 4 | 7 | 5.50 |
| Columbia | 4 | 4 | 8 | 5.75 |
| Cornell | 4 | 4 | 9 | 6.75 |
| Dartmouth | 1 | 4 | 4 | 4.00 |
| Harvard | 8 | 4 | 8 | 6.63 |
| Princeton | 2 | 4 | 5 | 4.50 |
| Cambridge | 5 | 5 | 10 | 6.80 |
| Oxford | 2 | 6 | 10 | 8.00 |
| Penn | 6 | 4 | 9 | 5.67 |
| Yale | 6 | 6 | 8 | 6.50 |
| Total | 40 | 4 | 10 | 6.25 |

Though the performance of our quiz-takers is poor, it is still better than random guessing. The latter would produce a symmetric distribution of scores centered at five correct answers, and the average score of 50%. Does this mean that there is a perceptible difference in quality between the music of Mozart and Salieri? Not necessarily so. From the feedback, I know that many quiz takers had previously heard some or all of Mozart's music used in the test. For example, one of the respondents wrote:

> I got 100%! I didn't expect to. I think I only got that score because I recognized all of the Mozart pieces.

Some musicians even know every note that Mozart wrote for their instrument, as the following note reveals:

> And when I heard the use of clarinets in one particular aria, I spotted that as Salieri not because I am smart and can distinguish between the two, but rather I know every note that Mozart wrote for clarinet, and that wasn't one of them.

Apart from explicit mentions in feedback, the fact that many quiz takers had previously heard some of Mozart's music is evident from the distribution of test scores. An average clip was identified correctly in 61.0% of the cases. An interesting thing is that this splits unevenly between Mozart and Salieri. An average Mozart clip was correctly identified in $p_m$ = 66.7% of the cases, while an average Salieri was correctly identified in only $p_s$ = 55.3%. The standard errors of both $p_m$ and $p_s$ are about 0.2%, so the difference $p_m - p_s$ = 11.4% is statistically significant. The obvious explanation of this observation is that some of the quiz-takers had previously heard some of Mozart music. If the test taker had previously heard Mozart music used in the clip, he will select it as Mozart. If he did not, he will have to use other criteria, for example, whether the music contains too many notes. As our quiz-takers likely have not previously heard any of Salieri music, the percentage of correct identification of a Salieri should be equal to the percentage of the correct identification of a Mozart in the case that the quiz-taker has not heard it before. This we can use to estimate the fraction, $f$, of Mozart clips, that our quiz-takers had previously heard. We can split the probability to identify a Mozart correctly, $p_m$, in two terms. If the taker has heard the Mozart clip before (what happens with probability $f$), he identifies it correctly with probability 1. If he has not heard the music before (what happens with probability 1-$f$), he identifies it correctly with probability $p_s$. Thus: $p_m = f + (1 - f) p_s$. From this follows: $f = (p_m - p_s) / (1 - p_s) = 25.6\%$.

Although the results of the quiz are biased in favor of Mozart, I'll take them at face value to quantify the difference in quality between the music. Table 2 shows for each clip the fraction of quiz-takers who selected it as "Mozart". The top-rated clip was ticked "Mozart" by 81% of people, and the bottom-rated by only 31%. What does this say about the difference in intrinsic quality?

A hundred years ago psychologist F.M. Urban conducted his classic study of *just perceptible differences* [3]. He asked the subjects of his experiment to compare a hundred-gram weight with a set of different weights. When two weights were very close, the subject's judgment was poor. However, statistically, they perceived the lighter weight to be heavier in less than fifty percent of the cases. For example, they judged 92, 100, and 104 gram to be heavier than 100 gram, in 10%, 50%, and 84% of trials respectively. I defined the musical "weight" of a hypothetical music clip, which is selected as Mozart by 50% of quiz takers, as 100 musical grams. Afterward, I inferred the weights of the clips by interpolating Urban's data (see Figure 3).

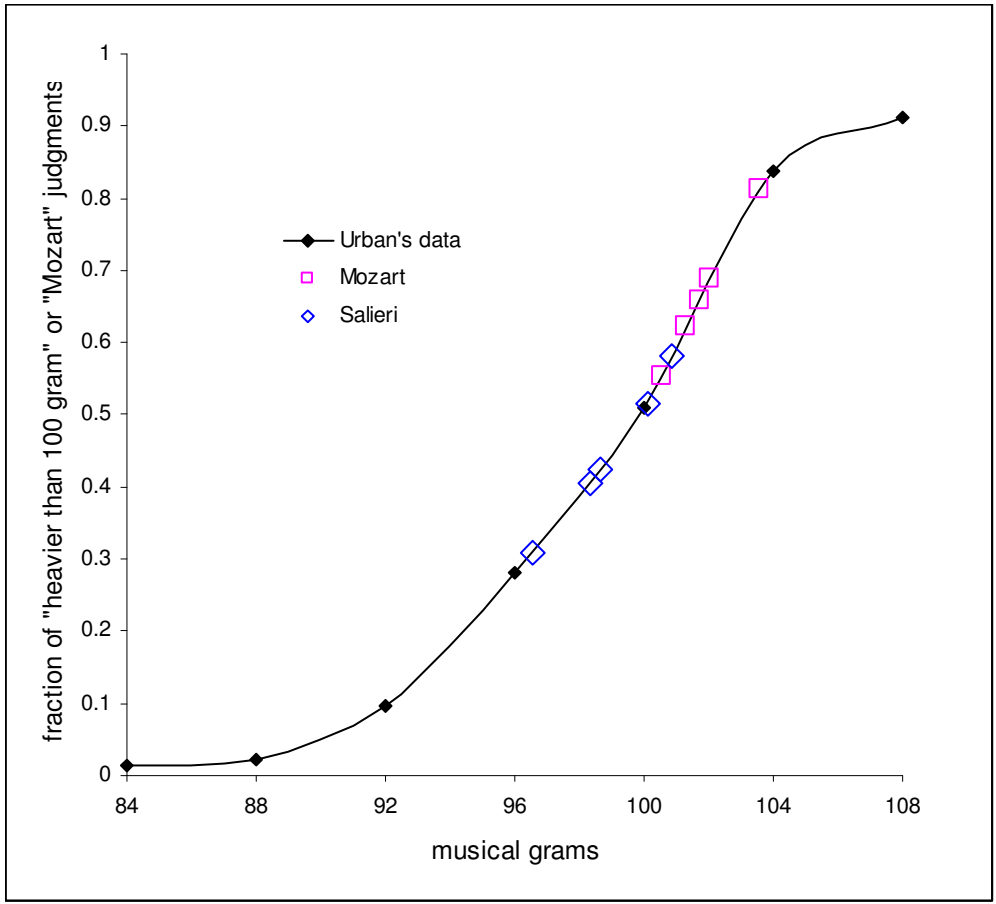

**Figure 2.** Small rhombs represent Urban's data on the fraction of "heavier" judgments for different weights (with the control weight of 100 gram). The line is the interpolation of that data. The large squares represent Mozart clips and the large rhombs represent Salieri. I adjusted musical weight so that the symbols fall on the interpolation line. You can listen to the clips on the quiz webpage [1].

**Table 2**. For each clip, the fraction of quiz takers, who selected it as Mozart, is shown alongside with clip's musical weight, determined by comparison with Urban's data. You can listen to the clips on the quiz's webpage [1].

| Clip number | Selected as Mozart | Musical gram | Composer | Opus |
|---|---|---|---|---|
| 9 | 81.2% | 103.6 | Mozart | Piano Concerto No. 9 E flat major (1777) *Allegro* |
| 6 | 68.9% | 102.0 | Mozart | La clemenza di Tito (1791) *Parto, parto* |
| 4 | 66.0% | 101.7 | Mozart | Le nozze di Figaro (1786) *Non so più* |
| 5 | 62.3% | 101.3 | Mozart | Don Giovanni (1787) *Vedrai, carino* |
| 7 | 58.1% | 100.9 | Salieri | Piano Concerto in C (1773) *Larghetto* |
| 10 | 55.3% | 100.6 | Mozart | Piano Concerto No. 18 B flat major (1784) *Allegro vivace* |
| 8 | 51.6% | 100.1 | Salieri | Piano Concerto in B flat (1773) *Allegro moderato* |
| 2 | 42.4% | 98.7 | Salieri | Palmira, Regina di Persia (1795) *Voi lusingateinvano* |
| 3 | 40.5% | 98.4 | Salieri | La Cifra (1789) *Alfin son sola ... Sola e mesta* |
| 1 | 30.8% | 96.6 | Salieri | La Secchia rapita (1772) *Son qual lacera tartana* |

The musical weights of the clips used in the test are given in the third column of Table 2. The difference in weight between the lightest (96.6 g) and the heaviest (103.6 g) clips is below ten percent. For comparison, in the sport of weight lifting men weighting between 94 and 105 kg belong to the same weight category [4]. I conclude that all peaces of music used in the quiz, when judged by their intrinsic qualities, fall into the *same weight category*.

The conclusion is not surprising since sportsmen are divided into weight categories so that they could compete with each other. A lightweight has no chance of winning a championship if he has to compete with heavyweights. However, an outcome of a match between two heavyweights is uncertain. And this is what we see in the quiz. Even the lowest ranking Salieri piece was selected as Mozart by 31% of quiz-takers. Therefore, what will be the next quiz-taker's judgment is highly uncertain.

A more straightforward comparison is with chess. In 1960[th,] Arpad Elo developed a rating system for chessplayers, which is now in common use [5]. Elo rating is a number computed based on player's performance, which we can use to compare players' relative strength. When the difference in rating between the players is 200 points, the expected score of the higher rated player is 77% of the number of games between these two players. When the difference is 400 or 600 the expected scores are 93% and 98% respectively. The players are divided into nine categories (see Table 3), from world championship contenders to novices, with the difference between bordering categories of 200 points.

While the aim of a chess player is to win a game, the aim of a composer is to win the ears of a listener. In the quiz, five Mozart pieces compete with five Salieri pieces. Let us take one particular quiz-taker, one particular Mozart's music piece and one particular Salieri's music piece. Since Mozart has a reputation of a genius and Salieri of a mediocricity, we assume that quiz-takers mark the music they like the best as Mozart. We will say that Mozart had won if the quiz-taker rated his piece as "Mozart" and Saliri's piece as "Salieri". If Mozart was rated as "Mozart" and Salieri also as "Mozart" we call it a draw. When both Mozart and Salieri are rated as "Salieri" is also a draw. When Mozart is rated as "Salieri" and Salieri as "Mozart" then Mozart had lost. Let us continue to stick to one particular quiz-taker, but now consider all music pieces. The number of Mozart's victories is equal to $M_c S_c$, where $M_c$ is the number of Mozart's pieces, identified correctly, and $S_c$ is the number of Salieri's pieces identified correctly. The number of draws is $M_c(5-S_c)+(5-M_c)S_c$. We will give 1 point for a victory and ½ point for a draw, like in chess. The expected Mozart's score is thus

$$M_c S_c + \frac{1}{2}\left(M_c(5-S_c)+(5-M_c)S_c\right) = \frac{5}{2}(M_c + S_c).$$

Since this score is a result of $5 \times 5 = 25$ "games", the average Mozart's score per game with Salieri is $\frac{1}{10}(M_c + S_c)$. And this is just the percentage of the correct answers given by our quiz-taker. Next, we need to average this over all quiz takers and we will get just the average test score. Thus the average Mozart's score in his competition with Salieri is 61%. According to the table in Chapter 2.11 of Elo's book [5] this corresponds to the rating difference of 80 points. Thus, Mozart and Salieri belong to the same player category. For comparison, consider the 1972 match between Fischer and Spassky who had the rating difference of 110. Fischer got 12.5 points in 20 games [5]. His average score per game was 62.5%.

**Table 3.** Elo rating of chessplayers.

| Elo rating | Player category |
|---|---|
| 2600 and up | World championship contenders |
| 2400 - 2600 | Grandmasters |
| 2200 - 2400 | National masters |
| 2000 - 2200 | Candidate masters |
| 1800 - 2000 | Amateurs – Class A |
| 1600 - 1800 | Amateurs – Class B |
| 1400 - 1600 | Amateurs – Class C |
| 1200 - 1400 | Amateurs – Class D |
| Below 1200 | Novices |

Several years ago, I conducted a similar experiment with modern art. And with similar results: immortal masterpieces turned out to be hardly distinguishable from my doodles [6]: the average test score was 66% correct. This puts me a 117 Elo points below abstract art grandmasters. About as much difference as between Fischer and Spassky. Therefore, according to Table 3, I also must be a grandmaster of abstract art or at least a national master.

In another test [7], I asked the takers to tell abstract art masterpieces from pictures painted by an ape. The average score earned by over 164 thousand people is 79%. A better result than on the other quiz, but one would expect so: the pictures were painted by members of different species. It is interesting that mistakes are at all possible. Apes are lagging behind Apestract Art grandmasters 230 Elo points. This means (see Table 3) that they should be national masters or at least candidate masters.

The results of art quizzes and Mozart-Salieri quiz are similar, but their meaning is different. After all, Salieri was not a Bremen musician. Thus, Mozart did not get in the same category with an animal. And even not in the same category with an arbitrary human. I did not compose the music for the quiz myself. I used the music of Salieri. I also could not perform the soprano arias myself and had to use the disc by Cecilia Bartoli. Thus, while great abstract artist is just as good as anybody, Mozart only appears overrated when compared to a small group of talented people.

My conclusion that Mozart is not that much better than Salieri may be shocking to some people. However, the test results are not the only evidence in its support. For example, Stendhal wrote in his autobiography [8]:

> I shall admit that for me only two composers have written songs of a perfect beauty: Cimarosa and Mozart, and I'd sooner be hanged rather than have to say honestly which of the two I prefer.

Stendhal gives the same weight as to Mozart to a composer, so unknown that Microsoft Word offered to replace his name with "Comatose." However in his lifetime Cimarosa was more popular than Mozart (though less popular than Salieri) as is evident from Table 4. Mozart was merely on the 7[th] place according to his popularity, trailing well behind Salieri who was on the second place after Paisiello. Similar rankings we observe when compare popularity of separate operas (see Table 5). Here Mozart's *Le nozze di Figaro* and *Die Entführung aus dem Serail* are on 9-11[th] places with 38 performances each. Trailing well behind Salieri's *Axur, re d'Ormus*, which endured 50 performances.

**Table 4.** Most popular composers in Mozart's Vienna, ranked by the number of performances of their operas. Note that Platoff [10], unlike Branscombe [9] counted only Italian operas. Thus, for example, the 38 performances of *Die Entführung aus dem Serail* (see Table 5) explains the major difference in their counts. Mozart is on the 7[th] place by either count.

| Composer | Opera performances, 1781-1791, according to Branscombe [9] | Italian opera performances, 1783-1792, according to Platoff [10] |
|---|---|---|
| Paisiello | 294 | 251 |
| Salieri | 185 | 167 |
| Martín y Soler | 141 | 140 |
| Cimarosa | 124 | 127 |
| Guglielmi | 112 | 112 |
| Sarti | 108 | 97 |
| Mozart | 105 | 63 |
| Gretry | 100 | |
| Dittersdorf | 72 | |
| Gluck | 70 | |
| Umlauf | 70 | |
| Anfossi | 56 | 51 |
| Storace | 40 | 41 |
| Benda | 37 | |
| Weigl | 33 | 27 |
| Monsigny | 33 | |
| Philidor | 21 | |
| Gazzaniga | 20 | 20 |

**Table 5.** Most popular operas in Mozart's Vienna.

| Opera | Composer | Performances Brascombe [11] | Platoff [10] |
|---|---|---|---|
| L'arbore di Diana | Martín y Soler | 66 | 65 |
| Fra i due litiganti il terzo gode (Im Trüben ist gut fischen) | Sarti | 63 | 58 |
| Il barbiere di Siviglia | Paisiello | 61 | 62 |
| Il re Teodoro in Venezia | Paisiello | 59 | 59 |
| Una cosa rara | Martín y Soler | 55 | 55 |
| Axur, re d'Ormus | Salieri | 50 | 51 |
| Die unvermuthete Zussamenkunft (La rencontre imprevue) | Gluck | 42 | |
| Gli astrologi immaginari (Die eingebildeten Philosophen) | Paisiello | 40 | |
| Zemire und Azor | Gretry | 38 | |
| Le nozze di Figaro | Mozart | 38 | 38 |
| Die Entführung aus dem Serail | Mozart | 38 | |
| La pastorella nobile | Guglielmi | 36 | 38 |
| Doktor und Apotheker | Dittersdorf | 33 | |
| Das Irrlicht | Umlauf | 32 | |
| La malinara | Paisiello | 31 | 32 |
| Gli sposi malcontenti | Storace | 29 | 29 |
| La bella pescatrice | Guglielmi | 27 | |
| La scuola de' gelosi | Salieri | 27 | |
| La grotta di Trofonio | Salieri | 26 | |
| Le gellosie villane | Sarti | 25 | |
| Il pazzo per forza | Weigl | 24 | |
| Il falegname | Cimarosa | 23 | |
| Le vicende d'amore | Guglielmi | 21 | |
| La contadina di spirito | Paisiello | 21 | |
| talismanoll talismano | Salieri | 21 | |
| Medea | Benda | 20 | |
| Le trame deluse | Cimarosa | 20 | |
| Il Burbero di buon cuore | Martín y Soler | 20 | |

Apart from described in this article online competition between Mozart and Salieri, there was a real life contest between the composers [11], [12]. One day in 1786 in Emperors palace two operas were performed. One was *Prima la musica, poi le parole* by Salieri and the other was *Der Schauspieldirektor* by Mozart. Josef II commissioned them both for the visit of his royal guests. Salieri's opera was a great success, while Mozart's had failed.

Mozart's contest with Muzio Clementi that took place in 1781 turned out not so bad. It resulted in a draw. Each composer performed on a clavier selections from his own compositions. Clementi performed his B-Flat Major sonata (op. 24, no. 2). Ten years later Mozart used the main theme of that sonata in the overture to *Die Zauberflöte* without any mentioning of Clementi [11], [12].

This is not the only case when Mozart borrowed from fellow composers. Cherubino's aria "Voi che sapete" in *Le nozze di Figaro* is based on the aria of Count Almaviva "Saper bramate" in Paisiello's *Il barbiere di Siviglia* [11], [12]. This, however, is not plagiarism, rather an ironical allusion to then famous opera. A more important thing is summarized by musicologist John Platoff in this words [10]:

Today, of course, the operas of Mozart's rivals are virtually never heard; if they were, it would be clear that much of what we think of as "Mozartian" is actually the general operatic style of the period.

When our quiz takers hear Salieri's music, they get their ears opened. "I think that Salieri is underrated" – wrote one of them. "I'll definitely be buying the Salieri Album" – wrote another.

**The notion that Mozart was a titan dwarfing his fellow composers is a superstition of our age**.